\documentstyle[11pt,aaspp4]{article}

\def\be{\begin{equation}}
\def\en{\end{equation}}

\def\msun{M_{\sun}}

\def\lsun{L_{\sun}}
\def\msunyr{M_{\sun} yr^{-1}}

\def\mdot{\dot{M}}

\begin{document}

\title
{THE COMPLEX PROTOSTELLAR SOURCE IRAS 04325+2402$^{1}$}

\author{Lee Hartmann\altaffilmark{2}, Nuria Calvet\altaffilmark{2,3}, Lori Allen\altaffilmark{2}, Hua Chen\altaffilmark{4}, 
and Ray Jayawardhana\altaffilmark{2}}
\affil{Harvard-Smithsonian Center for Astrophysics, 60 Garden St., MS 42,
Cambridge, MA 02138; hartmann@cfa.harvard.edu}

\altaffiltext{1}{Based on observations with the NASA/ESA Hubble Space Telescope,
obtained at the Space Telescope Science Institute, which is operated by
the Association of Universities for Research in Astronomy, Inc.,
under NASA contract No. NAS5-26555.}

\altaffiltext{2}{Visiting astronomer at the NASA Infrared Telescope Facility, which is operated
by the University of Hawaii under contract to the National Aeronautics and Space Administration.}

\altaffiltext{3}{also Centro de Investigaciones de Astronom\'ia, M\'erida, Venezuela}

\altaffiltext{4}{Currently at N.E.T., 6500 Paseo Padre Parkway, Fremont, CA 94555}

\begin{abstract}
We report {\it Hubble Space Telescope} near-infrared NICMOS observations 
of a remarkable low-luminosity Class I (protostellar) source in the Taurus
Molecular Cloud.  IRAS 04325+2402 exhibits a complex bipolar
scattered light nebula.  The central continuum source is resolved and may be multiple,
or may be crossed by a small dust lane.  Complex arcs seen in scattered light surround
the central source; the physical nature of these structures is not clear, but
they may reflect perturbations from multiple stellar sources or from time-dependent
mass ejection. 
A second, resolved continuum source is found at a projected
distance of approximately 1150 AU from
the central region, near the edge of a nebular lobe probably produced by outflow.
The images indicate that this second source is another low-luminosity young 
stellar object, seen nearly edge-on through a dusty disk 
and envelope system with disk diameter $\sim 60$~AU.
We suggest that the scattered light ``streaks'' associated
with this second source are limb-brightened outflow cavities in the dusty envelope,
possibly perturbed by interaction with the outflow lobes of the main source.
The nature of the companion is uncertain, since it is observed mostly in
scattered light, but is most probably a very low mass star or brown dwarf,
with a minimum luminosity of $\sim 10^{-2} \lsun$.
Our results show that protostellar sources may have multiple centers of infall
and non-aligned disks and outflows, even on relatively small scales.
\end{abstract}
\keywords{circumstellar matter, stars: formation, stars:pre-main sequence}

\section{INTRODUCTION}

Understanding the manner in which protostellar envelopes collapse to stellar
and disk systems is an essential problem in star formation.  While progress
has been made in testing the predictions of the standard theory of protostellar collapse  
using overall spectral energy distributions (SEDs) and source statistics of protostellar (Class I) 
sources in the Taurus molecular complex (e.g., Adams, Lada, \& Shu 1987; Butner et al. 1991; 
Kenyon, Calvet, \& Hartmann 1993 $=$ KCH; Kenyon et al. 1990, 1994), many details 
remain to be understood.  In particular, the roles of outflows and non-spherical initial
cloud structure in shaping protostellar clouds (Whitney \& Hartmann 1993; Kenyon et al. 1993;
Calvet et al. 1994; Hartmann et al. 1994, 1996; Li \& Shu 1996, 1997), 
and the way in which these effects
might or might not combine with the angular momentum distribution of protostellar
clouds to produce collapse to binary systems and disks (cf. Terebey, Shu, \& Cassen 1984;
Boss 1995, and references therein), is far from clear. 

Near-infrared imaging observations can play an important role in help resolving these problems.
While near-IR images are sensitive to very small amounts of material, due to
the efficiency of dust scattering and absorption,
such observations can be crucial in indicating the positions
and properties of the central stars, and in tracing envelope structure on
small scales difficult to observe in other spectral regions.

Here we report {\it Hubble Space Telescope} near-infrared imaging observations of a Taurus protostellar
(Class I) source, IRAS 04325+2402 (Beichman et al. 1986). This is a low-luminosity source
($\log L/\lsun = -0.14$) with other properties typical of Class I Taurus sources (KCH).  
This source is multiple, and exhibits a complex and confusing morphology, illustrating
the limits of SED analysis for understanding protostellar sources.

\section{OBSERVATIONS AND DATA REDUCTION}

Near-infrared observations of IRAS 04325+2402 
were obtained with the {\it HST} NICMOS Camera 2 
on 11/11/97 UT. Images were obtained with the F110W,
F160W, F205W, and F212N filters, the latter to search for H$_2$ emission.  
A five-position dither pattern with a separation of approximately
6 arcsec was used for the wide filters.  
Because the source is large and was not well-centered, only
four of the 5 dithered images were useful in producing the final images.
Dark sky images for F205W were obtained with a single chop position offset
by 40 arcsec.  Exposure times were 128 sec per frame in F205W and 40 sec in F160W.
The 80 sec F212N exposures provided no hint of excess H$_2$ emission, 
and the 40 sec F110W exposures showed only very faint nebulosity, 
so we do not consider them further.

The data were reduced using the standard ``calnica'' routine
in the STSDAS packages with updated calibration files and other tasks within 
IRAF\footnote{Distributed by NOAO, which is operated by AURA, Inc. under 
cooperative agreement with the National Science Foundation.}.
Offsets for the construction of the final image were
derived by cross-correlating image features.
Final images were created in two ways.  First, we translated the images
using fractional pixel shifts using the IRAF ``imshift'' task, and constructed
a median combined image using the ``imcombine'' task.
Second, images were combined using the ``drizzle'' task in IRAF/STSDAS 
(Fruchter \& Hook 1999). ``Drizzle'' employs a variable-pixel 
linear reconstruction algorithm, for the re-sampling of undersampled, 
dithered data, while correcting for geometric distortions and 
preserving photometry. In this method,  
pixels are mapped from an input image, or group of input images, 
into pixels in a subsampled output image. 
In this application, the input pixels were shrunk to 90\% of their
original size ({\bf pixfrac=0.9}) before being drizzled onto a 
grid with pixels half the original pixel size ({\bf pixscale=0.5}).
Weighted image masks, taking into account bad pixels and the coronographic 
spot, were prepared for each image and used in the drizzle. 
The images resulting from these two procedures are extremely similar.  For
basic image display we show the drizzled versions; for comparison with
models (\S 4), we use the imcombined images to maintain the original pixel scale
of the observations.

Supporting images at J, H, K, and an H$_2$ $2.12\mu$m filter  were obtained
with NSFCAM on the NASA IRTF telescope on 26 and 27 Nov 1998.
Data were obtained on two pixel scales, 0.15 and 0.3 arcsec/pixel,
on successive nights.  The data were reduced with standard IRAF routines.

\section{RESULTS}

\subsection{Images}

Figure 1 shows the large-scale K-band nebulosity as observed at IRTF with NSFCAM.
The large-scale bipolar structure is evident, consistent with typical
scattered light structures produced by bipolar outflows in young stellar
objects.   The northern scattered light lobe
is much brighter (and perhaps more conical) than the southern lobe,
and exhibits two bright structures toward its apex,  
in agreement with the K' image presented by Hodapp (1994).
An anomalous bright ``knot'', which we refer to as object C,
is apparent near the western edge of the northern lobe
(as also shown in the image of Hodapp), and is crossed by a curved bright 
streak proceeding from SE to NW which apparently passes through the
knot.  The streaks exhibit a peculiar ``corkscrew'' appearance; there
may be some corresponding structure, though not symmetrically placed,
in the southern lobe.  A comparison with the narrow H$_2$ filter images
shows that there is no measurable excess relative to either the broad-band K 
filter or a neighboring continuum filter, consistent with the F212N images.
Thus, the spectrum of C is not dominated by H$_2$ emission, and must be mostly continuum. 

The NICMOS F205W composite drizzled image is shown in Figure 2.
To preserve resolution the image has not been rotated from the observed
position angle of 38$^{\circ}$ for the y axis. 
The bright central region of the previous image is now resolved into
a bright, extended central object, which we call A/B.
Immediately S of A/B there is a bright ring or arc; to the N there is a 
peculiar $\Sigma$-shaped scattered light nebulosity.
Object A/B is not a point source; as shown in the expanded view in Figure 3a, there is
a double structure, possibly indicating two sources, or possibly indicating
a dark absorption lane running roughly EW across the object.

Figure 3b shows an enlargement of the F205W image near object C.  The object is
clearly resolved into a double-lobed emission separated by a dark lane.  This
structure is present in the three individual dithered images in which C appears.
This image structure is reminiscent of the HST WFPC2 images of HH30 
(Burrows et al. 1996) and of the faint companion to HK Tau (Stapelfeldt et al. 1998),
which have been interpreted as scattered light from the upper surfaces of
edge-on, flared disks around T Tauri stars, possibly augmented by
a dusty, relatively tenuous, envelope (Wood et al. 1998).

Figure 4 is a false-color rendition of IRAS 04325+2402 made by combining the
F205W (red) and the F160W (green) images. 
The W portion of the knot is relatively brighter in F205W than F160W, 
as indicated by the color scheme, suggesting greater extinction as would be expected
in a disk/envelope model (\S 4).   In other respects the F160W image closely 
parallels the F205W image structure, but with substantially lower signal-to-noise.

Table 1 gives positions of components A/B and C.  
The position given for A/B is actually the position of A,
the brightest portion (S component) of the extended object (Figure 3a);
component B is approximatley 0.074 arcsec east and 0.23 arcsec north of A.

\subsection{Photometry}

Determination of fluxes and conversion to magnitude systems 
is complicated by the difficulty of subtracting spatially-varying background,
by the very red colors of these objects, and by the non-standard
nature of the NICMOS filters.  
For the NICMOS data we used aperture corrections derived from observations of a very red standard
star (Oph S-1) kindly provided by Marcia Rieke; these aperture corrections agreed
well with results from the Tiny Tim program (Krist \& Hook 1997).
The fluxes reported in Table 1 use the standard calibration
without any attempt to correct for color.  Usually color terms are
ignored in determining fluxes and magnitudes, as we do here,
but we note that color corrections could be significant for such red objects.  
The corresponding magnitudes reported in Table 2 adopt the Vega zero point values. 

At IRTF, we measured a K magnitude (in the UKIRT system) 
of approximately 11.23 $\pm 0.05$ in a 7.2 arcsec diameter aperture 
centered on the main source (Table 2).
This is consistent with the K~=~11.27 reported by Kenyon et al. (1993) 
in an 8 arcsec diameter aperture;
however, we measured an (H-K)~=~1.79 (UKIRT) instead of the (H-K)~=~2.05
reported by KGWH.  Our color results must be viewed with some caution,
however, because there is a large color term in the transformation
of (H-K) between the IRTF system and the UKIRT or CIT systems 
((H-K)[IRTF]~$\sim$~0.8~(H-K)~[UKIRT, CIT]).  This color transformation has been
derived for stars which are much bluer than the red systems studied here,  
and so may not be appropriate to apply to IRAS 04325+2402.

The brightness of object C is uncertain for reasons of the difficulty in background
subtraction.  The F205W magnitude reported in Table 2 is considerably fainter than
the K magnitude from IRTF, but the difference can probably be understood due to
the difference in central wavelengths of the two filters, and the difficulty of
subtracting out the nebulous background at low spatial resolution.
(We used a larger aperture for the IRTF measurement because of seeing limitations.)
 
\section{DISCUSSION}

\subsection{Object C - nebula}

The image of Object C exhibits similarities to other {\it HST} images of young
stellar objects surrounded by circumstellar material in the form of disks and/or
envelopes, such as HH30 (e.g., Burrows et al. 1996).
To investigate this possibility further, we constructed a series of models
of disks illuminated by the central star, with and without
infalling envelopes.  Given the limited observational
information available (for example, the luminosity and mass
of the central star are unknown), and the number of parameters
needed for a disk plus envelope system, no unique model can be constructed.
Instead we adopt properties and parameters typical of models
of Class I sources and T Tauri disks to investigate the plausibility
of our overall interpretation.

The envelope is taken to have the density
structure of the standard protostellar infall model of
Terebey, Shu, and Cassen (1984).  The 
radiative equilibrium temperature of the envelope is calculated
as in KCH and Calvet etal. (1994). The parameters that
describe the envelope are ${\rho}_1$, 
the density the envelope would have at 1 AU if it were purely infalling
radially (i.e., if the material had no angular momentum), 
and the centrifugal radius $R_c$. 
The density parameter is proportional to the mass infall rate $\mdot$, 
$\rho_1 = 5.3 \times 10^{-14} {\rm g \, cm^{-3}} (\mdot / 10^{-5} \msunyr ) (M_*/\msun)^{-1/2}$.
The disk is assumed to be physically thin and vertically isothermal, with 
a radial temperature distribution 
$T = 10 K (100 {\rm AU}/R)^{1/2}$, where R is the cylindrical radius,
and surface density distribution
$\Sigma = {\Sigma}_0 (R_*/R)$, where
$R_*$ is the  stellar radius, as suggested by the detailed
models of D'Alessio et al. (1998, 1999). The disk has a fixed outer radius 
given by $R_c$, and its density distribution along the z axis at fixed
radius is determined from the equation of hydrostatic equilibrium.

To calculate an image of the system for a given
inclination to the line of sight, we solve
the transfer equation for the specific intensity
along rays that cross the system
at fixed projected coordinates on the plane of the sky.
The emissivity at each point along the ray is given 
by the sum of two terms (1) direct stellar light, attenuated by the intervening
material between the star and the point and scattered 
towards the observer with a Henyey and Greenstein phase function
($g = 0.03$) (cf. Calvet et al. 1992),
and (2) isotropic scattering towards the observer
of the diffuse radiation field in the envelope. This diffuse
field is obtained directly from the solution of the
radial radiative transfer equations in the envelope
(cf. Calvet et al. 1994). The first term corresponds
to the single scattering approximation, which describes well
the behavior of the  photons scattered by the disk (Burrows et al. 1996).
The second term represents multiple scattering in the envelope,
in the spherically symmetric approximation. This approximation
becomes more accurate at large distances, where the envelope
becomes spherical. Draine and Lee (1984) dust properties
have been assumed. The theoretical specific intensity map
was binned into pixels of 0.038" x 0.038", i.e., half the original
image pixel scale and convolved with the {\it HST} PSF function 
derived from Tiny Tim (Krist \& Hook 1997).  After convolution,
the images were rebinned back to the original pixel scale to
compare with observations.

Our best envelope $+$ disk model for Object C is shown
in Figure 5 compared with the F205W image.
The disk has a radius of 30 AU, and it is immersed in a low density envelope characterized by
${\rho}_1 = 10^{-14} {\rm g \, cm^{-3}}$ and $R_c$ = 30 AU.  The system is viewed
at an inclination angle to the line of sight of 83$^{\circ}$.
The theoretical image contours agree reasonably well with the observed image
contours, though we cannot reproduce the observed non-axisymmetric features with
an axisymmetric model.

Our investigation indicates that neither pure disk models nor 
pure envelope models of the type considered
here can satisfactorily reproduce the observed image. 
Models which include disks alone (left panel in Figure 6) do not have
emission extending sufficiently far in the perpendicular direction
to the disk plane (i.e., the disk image is too ``flat'').  Conversely,
pure TSC infall models do not exhibit sufficient indication of a 
dark lane (middle panel of Figure 6).
The situation is reminiscent of that of models for HH30 by Wood et al. (1998),
in which the brightest structures were produced by a disk but the
faint, highly-extended scattered light structures were attributed to
an envelope.

The model calculations for F160W are in approximate agreement with the observations.
We do not attempt a detailed comparison because of the substantial noise in this image.
The model extinction is (slightly) higher toward the southeastern component, in 
qualitative agreement with the (modest) color difference observed.  
Assuming an intrinsic color of the central object of (H-K)$ = 0.6 \pm 0.3$ (see \S 4.2),
the model predicts (H-K)$ \sim 1.8 \pm 0.3$.
Given the uncertainties in the data and the unknown nature of the colors
of the central source, this agreement is sufficiently close to demonstrate
the general plausibility of the model.

We conclude that a model which includes both a disk and an envelope
can qualitatively account for the image of this object.
This model also may explain the scattered light ``streaks'' or structures
associated with C.  We suggest that these streaks might delineate the edges of an outflow
cavity produced by C (see Figure 7).  

If we assume a stellar mass of $0.1 \msun$ for C (\S 4.2), 
the implied infall rate for the envelope of our model is $\sim 6 \times 10^{-7} \msunyr$.
This is below the average infall rate of $\sim 4 \times 10^{-6} \msunyr$ estimated
for typical Taurus Class I sources, and lower than the overall
$\sim 1.3 \times 10^{-5} \msunyr ( M / 0.5 \msun )^{1/2}$
derived for the main source in IRAS 04325+2402 by KCH (though
substantially larger than the $3 \times 10^{-8} \msunyr$ 
adopted by Wood et al. [1998] for their HH30 model).  
From the observed source morphology
it appears plausible that C is encountering a lower-density environment than typical
of the regions near A/B, possibly as a result of orbital motion away from the main central mass.

\subsection{Object C - central star}

The nature of the central star in C is difficult to determine given the complications involved,
but we can make some estimates.  For this purpose we use the ground-based IRTF measurements 
because they are (approximately) on a standard magnitude system, though the lower
spatial resolution makes extraction of the source with respect to the nebular
background more difficult.
The very red (H-K)$= 1.95$ color of C indicates that the central star must
be heavily reddened.  The intrinsic colors of (M-type) Classical T Tauri stars in Taurus 
(young stars with inner circumstellar disks) range
from about (H-K)$\sim 0.3$ to $\sim 0.9$ (Meyer et al. 1997)
(for objects with no contribution from a dusty envelope).
Adopting an intrinsic color (H-K)$=0.6 \pm 0.3$, the excess is 
E(H-K)$=1.3 \pm 0.3$. Using A(K)~$=1.4$E(H-K) from Cohen et al. (1981), 
and ignoring the difference in colors between CIT and UKIRT systems,
we find A(K)$=1.8 \pm 0.4$, so K$_{\circ} = 12.8 \pm 0.4$.
Using the results of Wilking, Greene, \& Meyer (1999; their Figure 1) 
for evolutionary tracks, and adopting an age of $3 \times 10^5$ yr as typical of
Class I sources (Kenyon et al. 1990, 1994), the result is
a luminosity of $\sim 10^{-2} \lsun$ and
a mass of $\sim 0.03 - 0.02 \msun$, well into the brown dwarf range.

However, this estimate is almost certainly a lower limit to the stellar luminosity (and 
therefore to the stellar mass) 
because the object is undoubtedly seen entirely in scattered light, and thus the observed
colors underestimate the extinction directly toward the central star.  
For the model shown in Figure 6, the observed flux at K is approximately 4\% of the true
stellar flux.  Using this factor to correct the observed K magnitude would result in $K_{\circ} \sim 11$ and
a mass of about $0.06 \msun$ from the Wilking et al. (1999) estimates. 
Small changes in envelope parameters easily could increase the true K brightness by substantial factors.
At the other extreme, if we assume that C supplies {\it all} of the heating
responsible for the IRAS long-wavelength emission (cf. KCH), 
then the luminosity indicates a central star of $\sim 0.25 \msun$ 
from the evolutionary tracks of D'Antona \& Mazzitelli
(1997).  Given that the main source A/B seems to be responsible for most of the scattered light
in the main nebular lobe, this seems unlikely.  Thus 
the central star in C is probably either a very low mass star
or a brown dwarf; the F205W observations may provide
the first image of a protoplanetary disk around a brown dwarf.  
Infrared spectra should be obtained to better constrain the nature of C.

\subsection{Object A/B}

A/B lies near the apex of the main bipolar reflection nebula (Figure 1), and is probably
the main luminosity source in the system.  This geometrical relation
indicates that A/B is probably the source of bipolar ejection which produces the large-scale 
reflection nebula structure.  The alignment of these cavities are very different than
the assumed alignment of object C (see Figure 7).  Thus, it appears that even relatively
close, and probably coeval, multiple protostellar systems may have significantly
different outflow/disk orientations.

The large extinction toward A/B (as indicated by the colors, and by the
extreme faintness of the source at F110W and  J), indicates that A/B is also a center of
infall for the extended dusty envelope, as envisaged by the models of KCH and Kenyon et al. (1993b).
Since A/B is probably more luminous and thus possibly more massive than C,
it would not be surprising if most of the infall is directed toward it rather than toward
C.  The presence of multiple gravitating
centers in this system may account for some of the peculiar nebular structure.

Whether A/B is itself a multiple stellar system, or simply a complex scattered
light structure around a single star, is unclear.  While we do not offer a model
for the peculiar scattered light structures in the vicinity of this source, we note
that they appear to lie along portions of the outflow cones of the main source.  The scattered
light images may indicate complicated interactions of a structured infall region with outflow
(Figure 7), although the precise way this might happen is far from clear.  

The lower (southern) lower scattered-light arc is reminiscent of ground-based 
images of circumbinary dust rings around T Tauri stars in Taurus, as for example
is observed in GG Tau (Roddier et al. 1994; see also Dutrey, Guilloteau, \& Simon 1994)
and UY Aur (Close et al. 1998).  
The inner ring radii in these two objects are around 200 AU and 500 AU, respectively;
for object A, the diameter of the southern ring is about 1.3 arcsec $\sim$ 180 AU at 140 pc. 
In this picture, the ring could be produced by the tidal action of an orbiting 
binary or multiple protostellar system.  A possible problem with this picture is that
A/B does not seem to lie inside of the ring, but somewhat above it.

Object A/B is probably the main gravitating center toward which most of the dusty envelope
is presumably collapsing.  The model of KCH, which was adjusted to fit the sparsely-observed
long-wavelength SED of the object, suggested an infall rate of $\sim 1.3 \times 10^{-5} \msunyr
( M / 0.5 \msun )^{1/2}$.  Even with a large centrifugal radius, $R_c = 300$~AU,
to reduce extinction toward the main source, the model did not account for the observed near-infrared emission.
The attempt of Kenyon et al. (1993) to explain the near-infrared fluxes and large-scale image including
outflow cavities in the collapse model led them to estimate 
$\mdot \sim 4 - 6 \times 10^{-6} \msunyr ( M / 0.5 \msun )^{1/2}$
and $R_c \sim 50-200$~AU.  Given the complex scattered light structure seen in our images on
scales of 100 AU and less, it is clear that any estimate of the centrifugal radii of collapsing
envelopes from near-infrared fluxes alone is very uncertain.

\subsection{Overall structure}

The peculiar ``corkscrew'' structure
in the northern lobe of the main nebula may be the result of interacting flows of C and A.
In this connection we note that the observed (monopolar) CO outflow (Heyer et al 1987) in the region is
oriented to the northwest of IRAS 04325$+$2402.  With a position angle of about $-45^{\circ}$, 
the outflow is closer to the orientation expected for an outflow from C, and completely inconsistent
with the nearly north-south orientation of the main scattered light nebula (Figure 1).
However, the CO outflow extends over such a large scale that its connection with
IRAS 04325$+$2402 is unclear; moreover, the observed CO motions are redshifted, whereas one
might expect to observe a blueshifted lobe from object C (cf. Figure 7).

In many respects IRAS 04325$+$2402 resembles another Class I Taurus protostar with similar bolometric
luminosity and near-IR colors, IRAS 04361$+$2547 (KCH), which has
recently been imaged with NICMOS by Terebey et al. (1998).
IRAS 04361$+$2547 also shows a complex central structure, scattered light ``streaks'',
and a faint companion.  Terebey et al. interpret their complex central structure to
mean that the main source is a binary.  As mentioned above, this may apply to object A as well,
though we are hesitant to make a firm interpretation since the possibility also exists that these are simply
scattered light structures intersected by absorbing dust lanes.  

The point-source companion found by Terebey et al. in 04361$+$2547 also lies in a scattered light
structure, at a similar distance from the main region (1400 AU vs. the 1150 AU projected distance between 
objects A/B and C).  This companion has roughly similar colors to object C but is about a factor of 10
fainter.  Terebey et al. estimated the companion mass as 2-5$M_J$, and suggested that
its position at the end of a scattered light filament might be caused by
ejection from the central region over the last 1000 yr.
Our object C differs in that it is clearly too bright to be a $M \lesssim 10 M_J$ planet and
has its own solar-system sized disk.   Another difference is that the streaks or filaments 
do not point back to A/B, unlike the situation in 04361$+$2547.
If object C's disk had been a factor of two or so smaller, it would have appeared to be
a point source, raising the possibility that the Terebey et al. object might also appear 
anomalously faint for its color as a result of disk/envelope extinction. 

\section{Conclusions}

The Class I Taurus source IRAS 04325$+$2402 has been shown to be a multiple system, with at least
one additional protostar lying a projected distance of $\sim 1150$~AU from the central source,
and that central source may itself be multiple.
The NICMOS images of the companion are consistent with
the typical model of a low-mass protostar, in which
infalling material from a rotating envelope lands upon, and builds up, a disk.
The dust envelope structure around the main source appears more complex, but suggests an independent
center of collapse.  The orientation of the main bipolar cavities seen in scattered light, 
probably driven by outflow from the main source, lies at a substantial angle to the presumed
disk plane of the companion object.  This suggests that the orientation of disk systems can
be quite different even in relatively close multiple protostellar systems.

We wish to thank Bill Vacca for his very helpful support at IRTF, Karla Peterson and 
Louis Bergeron at STScI, and Marcia Rieke for providing standards. 
This work has been supported in part by
NASA through grant number GO-07413.01-96A from the Space Telescope
Science Institute.

\newpage

\begin{table*}[ht]
\begin{center}
\begin{tabular}[h]{ l c c c c r}
\multicolumn{6}{c}{ TABLE 1} \\
\multicolumn{6}{c}{ NICMOS POSITIONS AND PHOTOMETRY} \\
\tableline
Object & $\alpha$(J2000) & $\delta$(J2000) & F110W ($\mu$Jy) & F160W (mJy) & F205W (mJy) \\
\hline
A/B\tablenotemark{a}    & 4 35 35.37      & 24 08 19.5      & 93 $\pm 1$      &  3.1  $\pm 0.3$ & 14.4 $\pm 0.5$ \\
C      & 4 35 35.29      & 24 08 27.6      & $< 20$          &  0.14 $\pm 0.02$ & 0.47 $\pm 0.05$ \\
\hline
\end{tabular}
\end{center}
\tablenotetext{a}{Position given for A.  B is 0.07 arcsec E, 0.23 arcsec N of A.}
\end{table*}

\begin{table*}[hb]
\begin{center}
\begin{tabular}[h]{ l c c c c r}
\multicolumn{6}{c}{ TABLE 2} \\
\multicolumn{6}{c}{ IRTF and NICMOS MAGNITUDES} \\
\tableline
Object            &   K  & H-K  & m(F110W)\tablenotemark{a} & m(F160W)\tablenotemark{a} & m(F205W)\tablenotemark{a}\\
\hline
A/B\tablenotemark{b}& 11.23 $\pm 0.05$ & 1.79 $\pm 0.05$  & 18.3 $\pm 0.2$ & 13.89 $\pm 0.1$ & 11.80 $\pm 0.05$ \\
C\tablenotemark{c}& 14.61 $\pm 0.1$  & 1.95 $\pm 0.1$   & $> 20$         & 17.27 $\pm 0.2$ & 15.52 $\pm 0.15$ \\
\hline
\end{tabular}
\end{center}
\tablenotetext{a}{Magnitudes on HST Vega system.}
\tablenotetext{b}{Magnitudes in 7.2 arcsec diameter apertures.}
\tablenotetext{c}{Magnitudes in 1.37 arcsec diameter aperture (NICMOS) and 1.8 arcsec diameter
aperture (IRTF).}
\end{table*}

\newpage

\begin{figure}
\caption{K-band image of IRAS 04325+2402, obtained at IRTF.  North is at the
top and the displayed region is 76 arcsec on a side.}
\end{figure}

\begin{figure}
\caption{HST NICMOS F205W image mosaic of 04325+2402.  This ``drizzled'' Camera 2 image
(see text) is 15.6 arcsec on a side.  The orientation of this and the following HST
images is such that the vertical axis lies at a position angle 38 degrees east of north.}
\end{figure}

\begin{figure}
\caption{(a) Expanded view of the central region in Figure 2, 5.1 arcsec (horizontal)
by 4.2 arcsec (vertical).
(b) Expanded view of the companion from Figure 2, 4.56 arcsec by 3.42 arcsec.
The companion is clearly resolved, with a dark lane running roughly vertically
across the image (see text).  Note that the pixels in these drizzled images are one-half
the original pixel scale.}
\end{figure}

\begin{figure}
\caption{False color image of 04325+2402 composed of the F205W (red) and F160W 
(green) images.  One observes that the eastern portion of object C is redder
than the western lobe, in agreement with our model which predicts that the
fainter western lobe is extincted by a dusty envelope (see text).}
\end{figure}

\begin{figure}
\caption{Image and contour plot comparison of disk (left), envelope (center), and combined
disk and envelope (right) models for the F205W image of Object C.  Models have been
computed at the same pixel scale as the original, non-drizzled image (see text).  
Contour levels are at in linear steps of 20\% of the peak.}
\end{figure}

\begin{figure}
\caption{Contour plot comparison of F205W image (upper) with model disk and
envelope image from Figure 6 (right)(see text).  Here the comparison is made with
the median-combined dithered image at the original pixel scale of the observations.
Contour units are in linear steps of 10\% of the peak.}
\end{figure}

\begin{figure}
\caption{Schematic diagram of suggested source geometry (see text).  The regions exterior
to the cavities are filled with dusty gas, some of which is presumed to be falling in toward
the two mass centers A/B and C.}
\end{figure}

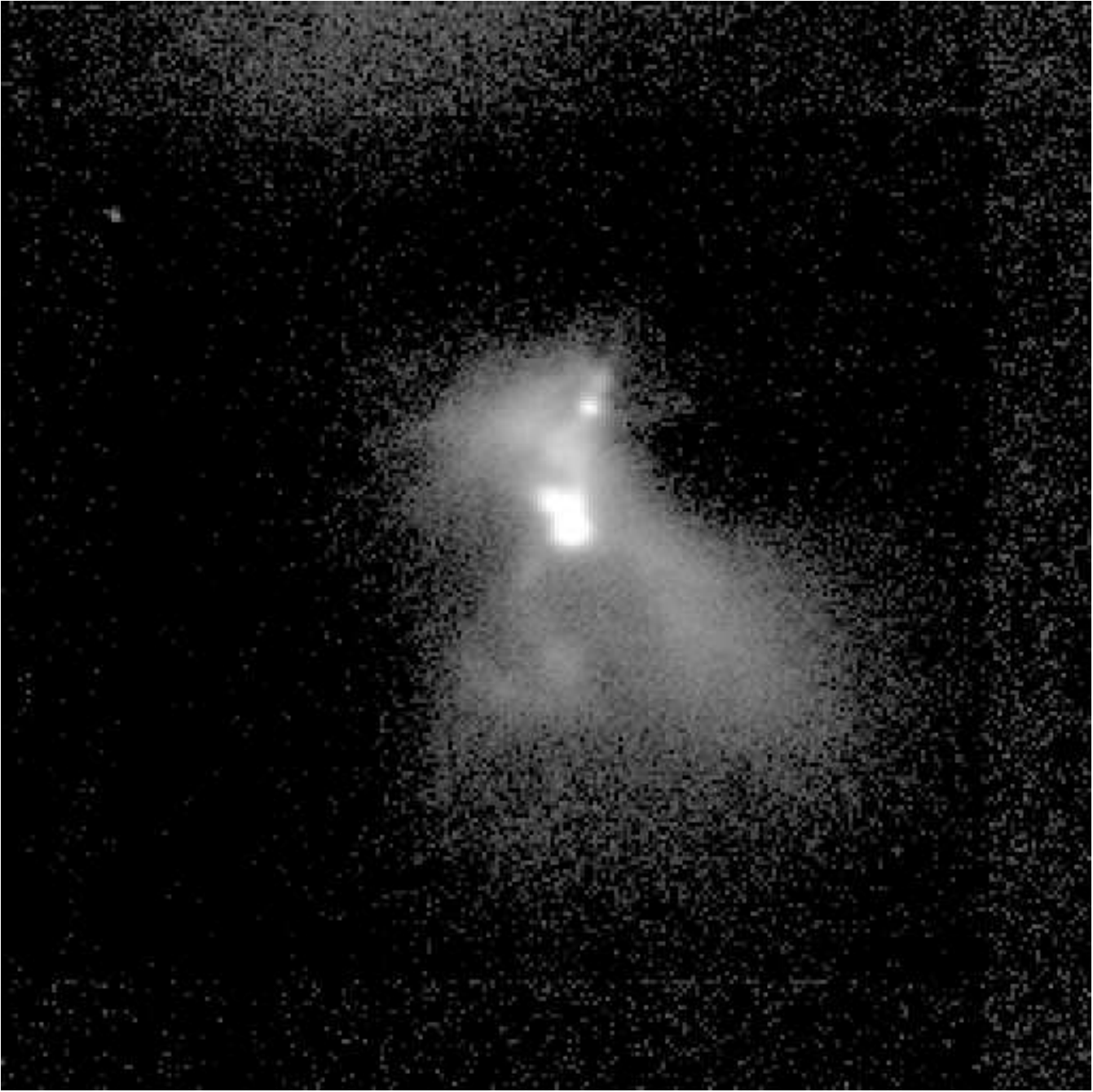
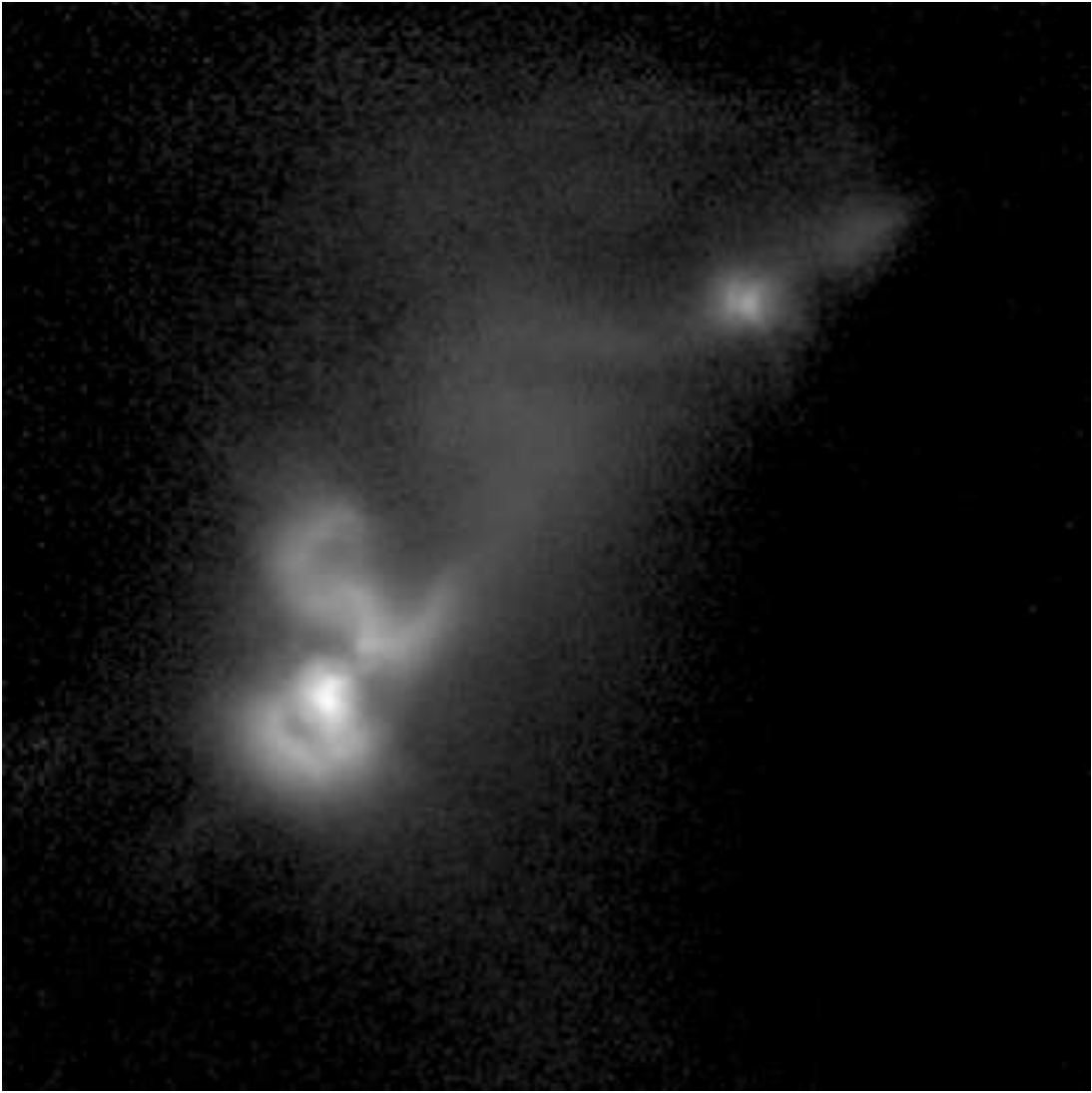
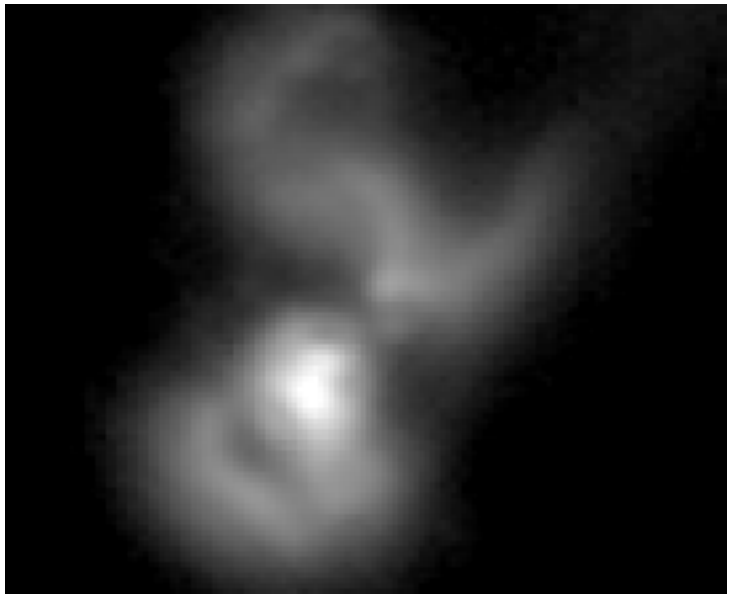
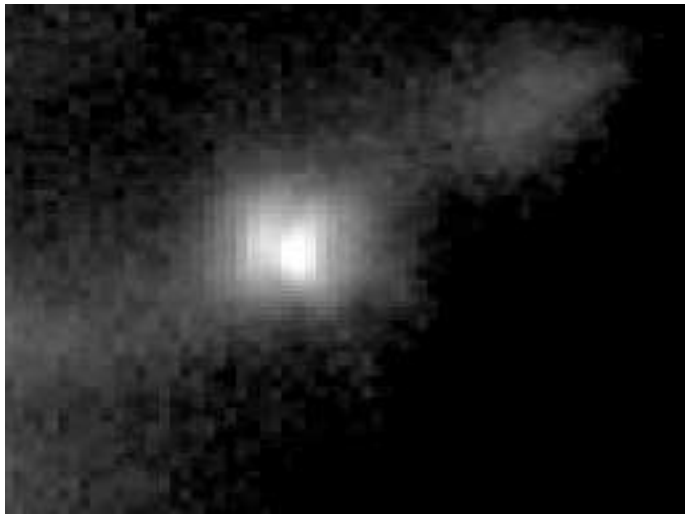
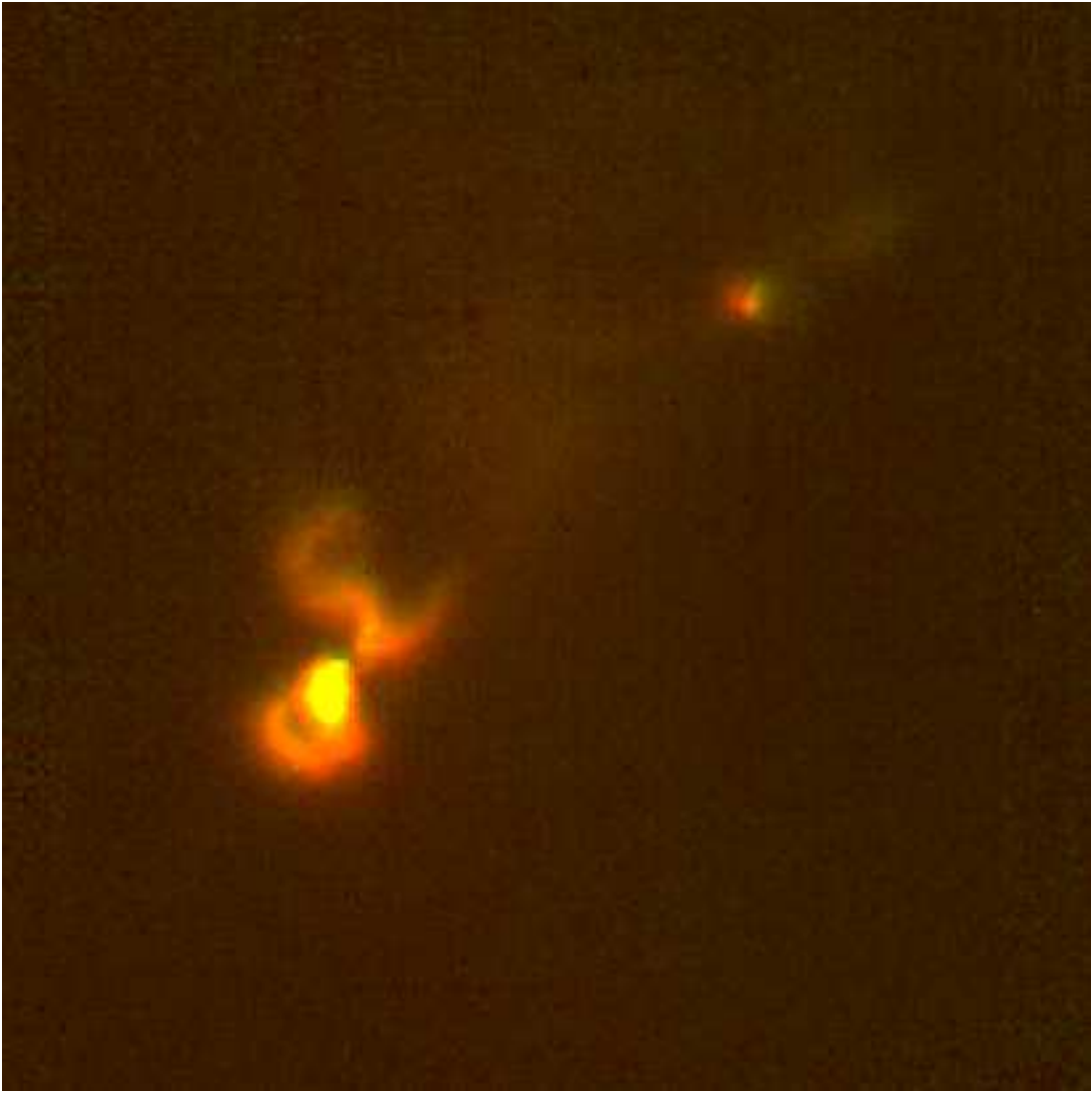
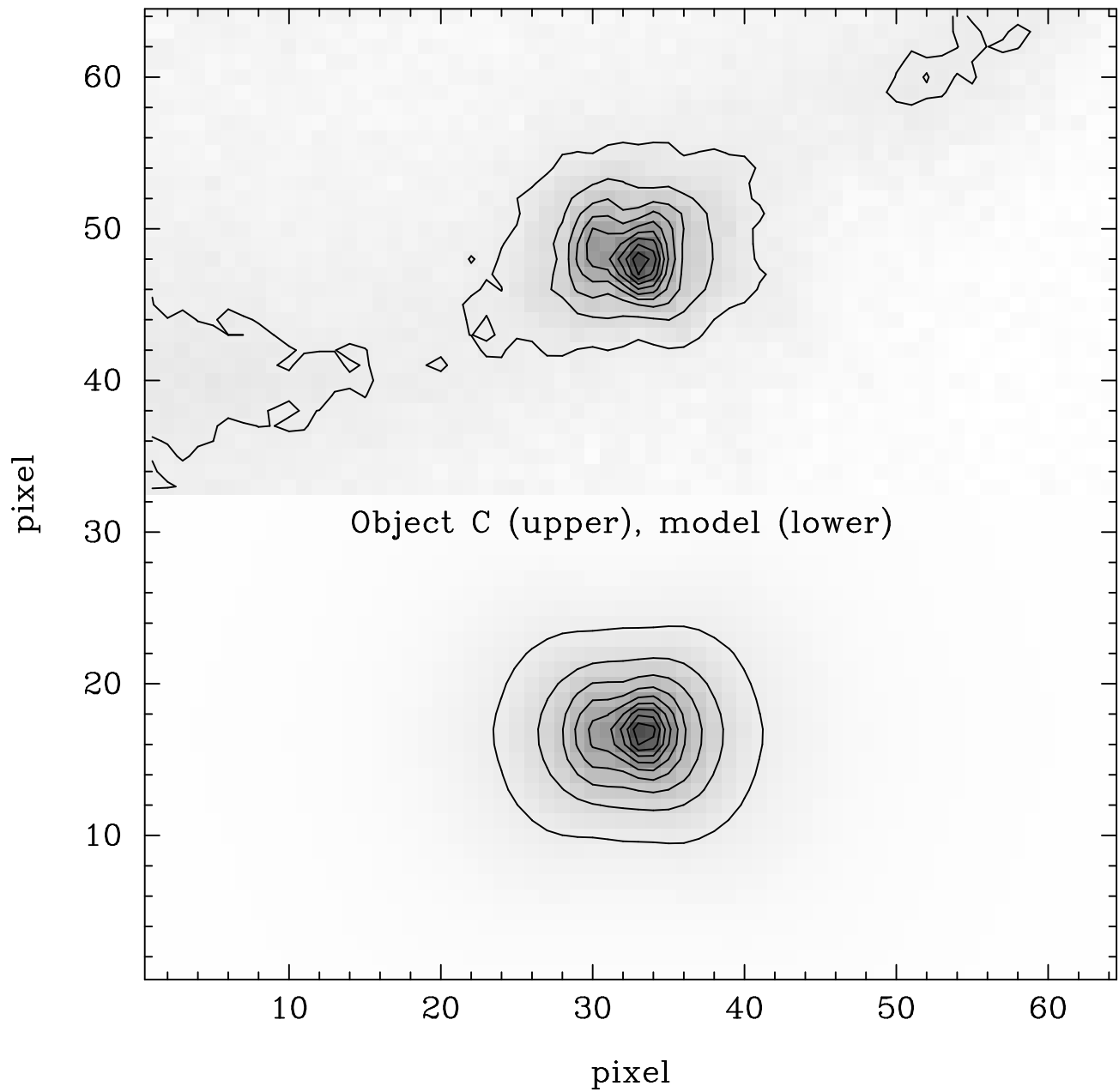
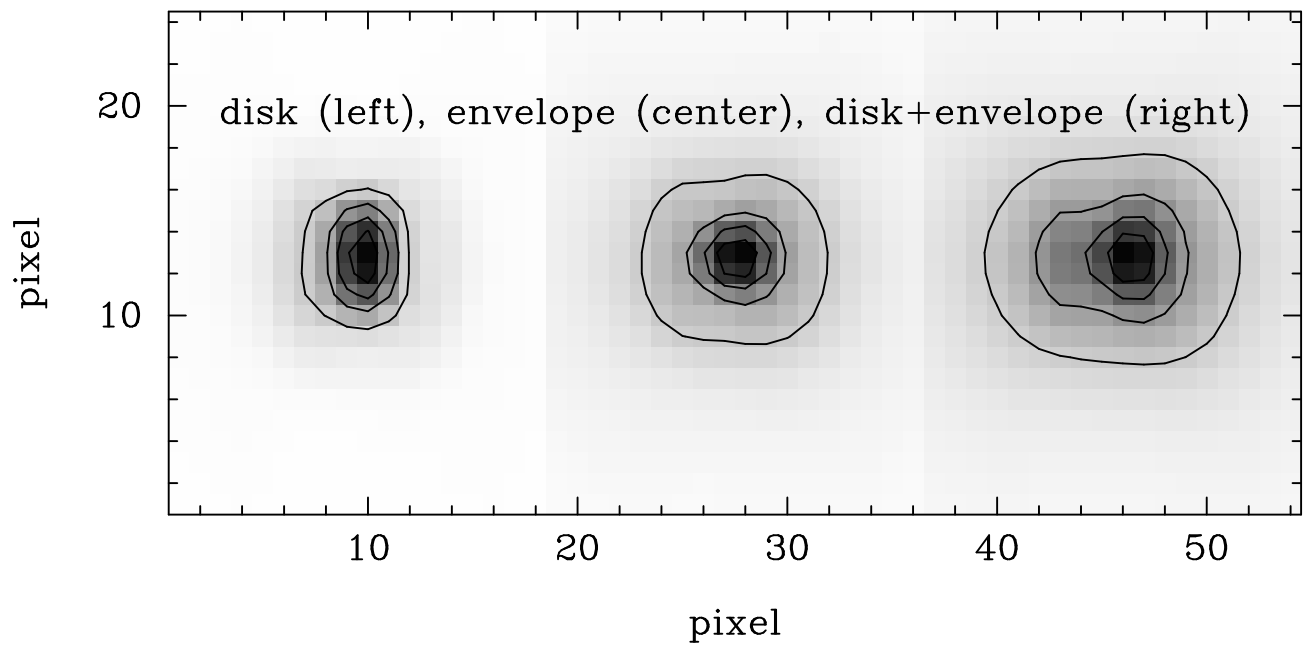
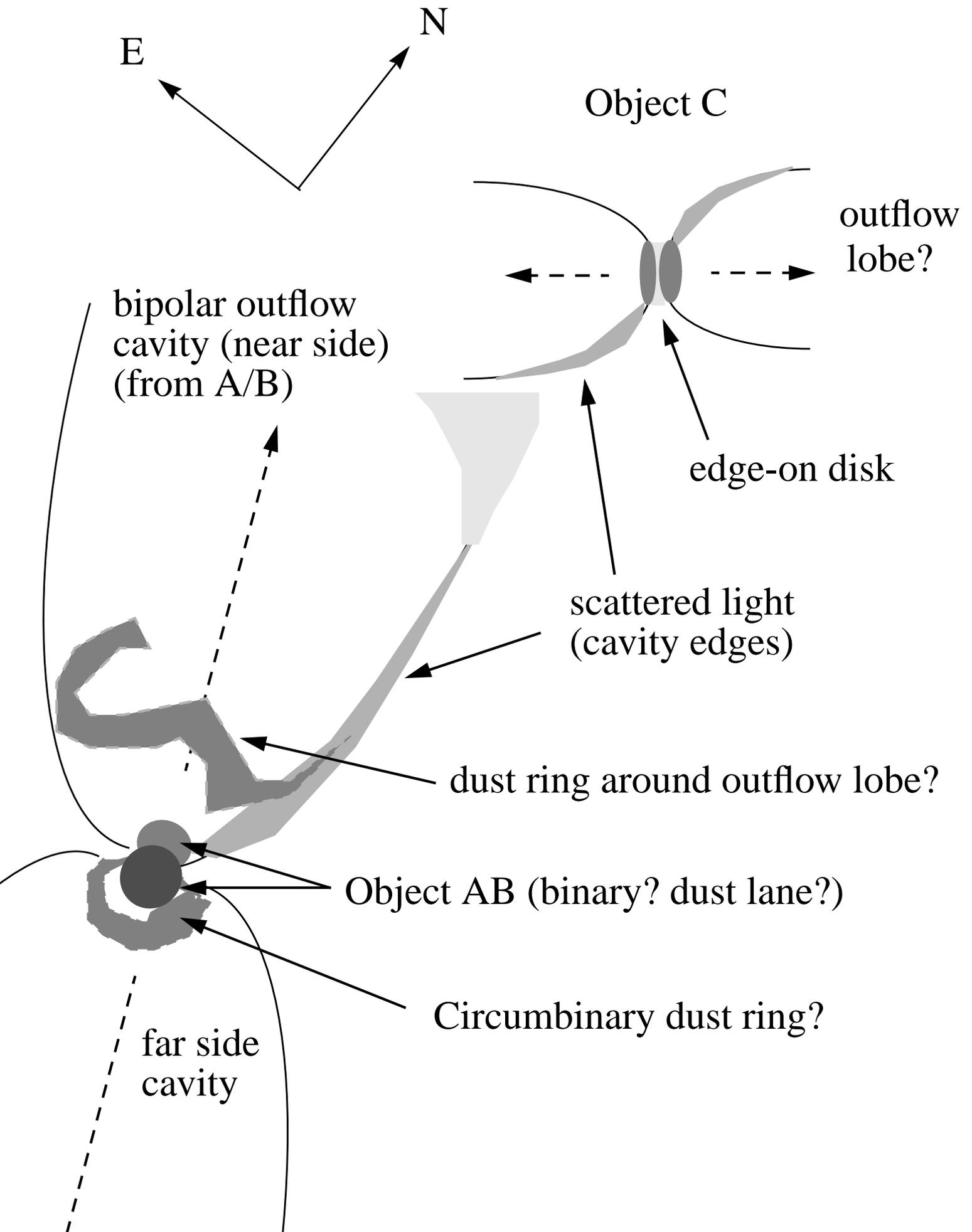
\end{document}